\documentclass[preprint,showpacs,preprintnumbers,amsmath,amssymb]{revtex4}

\usepackage{graphicx} \usepackage{dcolumn} \usepackage{bm}

\begin{document}
\title{Shallow Deep Transitions of Neutral and Charged Donor States in
Semiconductor     Quantum     Dots}     \author{R.     K.      Pandey}
\email{rajanp@iitk.ac.in} \author{Manoj  K.  Harbola, Vijay  A. Singh}
\affiliation{Physics  Department,  I.I.T.   Kanpur,  Uttar  Pradesh  -
208016, INDIA} \date{\today}

\begin{abstract}
We carry out  a detailed investigation of neutral  ($D^0$) and charged
($D^-$)   impurity  states  of   hydrogen-like  donors   in  spherical
semiconductor quantum  dots.  The investigation is  carried out within
the effective  mass theory  (EMT). We take  recourse to  local density
approximation (LDA)  and the  Harbola-Sahni (HS) schemes  for treating
many-body  effects.   We  experiment   with  a  variety  of  confining
potentials:  square, harmonic  and  triangular.  We  observe that  the
donor level  undergoes shallow  to deep transition  as the  dot radius
($R$) is reduced.   On further reduction of the  dot radius it becomes
shallow again.  We  term this non-monotonic behaviour \textbf{SHADES}.
This  suggests the  possibility  of carrier  {\textbf{\textit{``freeze
out''}}} for both  $D^0$ and $D^-$. Further, our  study of the optical
gaps also reveals a {\textbf{SHADES}} transition.
\end{abstract}

\maketitle

\section{Introduction}
\label{s:intro}

Impurities   play  a   central  role   in   semiconductor  technology.
Performance of a semiconductor  device is dictated by shallow dopants.
Shallow dopants  like phosphorous and boron alter  the conductivity of
bulk silicon by several orders of magnitude. Deep defects on the other
hand are known to degrade  device performance. An interesting point is
that the concentration  of these impurities is miniscule:  less than a
tenth  of a  percent.   Relevance of  the  role of  shallow donors  in
semiconductor   quantum  dots  (QDs),   which  are   essentially  zero
dimensional structures, thus cannot be overestimated.

Nearly  two  decades  ago  Bastard \cite{bast81}  reported  the  first
calculation  of  the  binding  energy  of  a  hydrogenic  impurity  in
two-dimensional  quantum  wells  (QW).  Bryant  \cite{brya84,  brya85}
extensively  studied  hydrogenic  impurity states  in  one-dimensional
quantum well  wires (QWW). Zhu \textit{et al.}   \cite{zhu90} were the
first to study the effect  of hydrogenic impurity in spherical quantum
dots  (QDs).  The  system  chosen  was ($GaAs  -  Ga_{1-x} Al_x  As$).
Calculations based  on the hydrogenic  impurity states in  presence of
electric  or magnetic  field have  also  been reported  over the  past
decade \cite{xiao96,  ribe97, core01}. Ranjan  and Singh \cite{ranj01}
reported studies  of neutral donor  states ($D^0$) in II-VI  and III-V
QDs. They examined the effects of a size dependent dielectric constant
$\epsilon (R)$, where $R$ is the  radius of the spherical QD.  As long
back as 1990 Pang and Louie \cite{pang90} discussed the negative donor
centers in  semiconductor quantum wells.  They  calculated the binding
energy of negative donor state  ($D^-$) in presence of magnetic field,
solving the  effective mass model  exactly by diffusion  quantum Monte
Carlo method.

Zhu  \textit{et al.}   \cite{zhu92, zhu94}  worked out  the  effect of
dimensionality and potential shape on  the binding energy of $D^0$ and
$D^-$ states of  the impurity.  Further, they also  studied the effect
of electron  correlation in the  binding energy. They showed  that the
shape  of  the confinement  is  less  important  than the  confinement
strength.  They also mentioned  the importance of electron correlation
in their  results.  Ranjan and  Singh \cite{ranj01} made  a systematic
study  of the non-monotonic  shallow deep  transition (SHADES)  of the
impurity in a semiconductor QD.   SHADES occurs when one decreases the
size of the QD.  First the  impurity level becomes deep and then below
a critical size it becomes shallow again.

In the present work we study both the neutral $D^0$ and the negatively
charged $D^-$ hydrogenic donor states in a three-dimensional spherical
quantum dot  (QD).  We  study the effect  of the shape  of confinement
potential on the binding energy of these donors. Many-body effects are
taken into account by employing the density functional methodology. To
treat the exchange effects accurately, particularly for negative ions,
we employ the Harbola-Sahni (HS) scheme \cite{harb89} for the exchange
part of the interaction. We also  compare the HS results with those of
the local density approximation (LDA).

\section{Model}
\label{s:model}
Our calculations have been performed  on a spherical QD We have solved
the Schroedinger equation  self-consistently within the effective mass
theory  (EMT)   approximation.  We  have   included  electron-electron
interaction effect  within the  local density functional  approach and
the Harbola-Sahni scheme.   The Hamiltonian of the system  is given by
the following expression
\begin{eqnarray}
H  & =  &  -  \sum_{i=1}^N \frac{1}{2  m^*}  \nabla_i^2 +  \frac{1}{2}
\sum_{i=1}^{N} \sum_{j \neq i}^N \frac{1}{\left| \vec{r_i} - \vec{r_j}
\right|} + V_{ext} (r) - \frac{Z}{r} \label{eq:mbody}
\end{eqnarray}

The first term  in the above Hamiltonian is the  kinetic energy of the
electrons,  the  second  term  is  the  electron-electron  interaction
energy, the third term is the external potential for N-electron system
and  the last  term is  the impurity  potential.  This  Hamiltonian is
written in Hartree's atomic unit, where $\hbar = m_e = e = 1$ and unit
of energy is 27.2 $eV$ and that  of length (size) is 0.53 \AA.  In the
above equation $m^*$ is the  effective mass of the electron inside the
QD in  units of  $m_e$, the free  electron mass.  The  above many-body
Hamiltonian  can be  reduced to  a single  particle  Schroedinger type
Hamiltonian for  the ground state within the  Kohn-Sham formulation of
density functional theory (DFT) \cite{kohn65}.

\begin{figure*}
\includegraphics{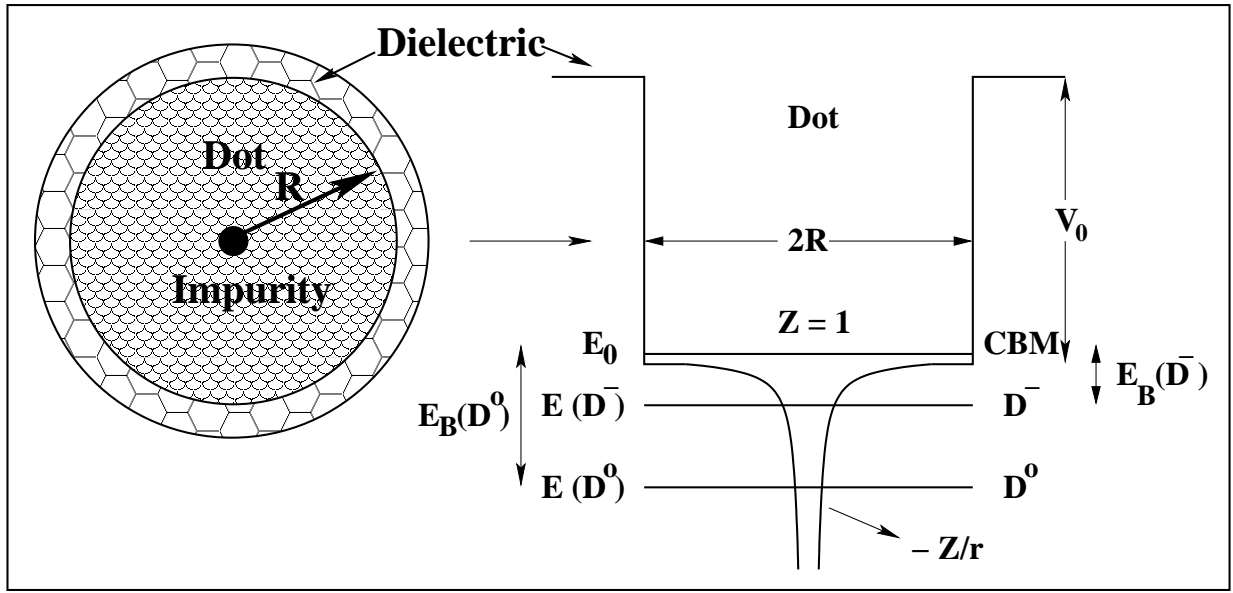}
\caption{\label{fig:msfig0}  A schematic  diagram, shows  the impurity
doped spherical quantum dot (QD).  The left side of the figure depicts
a  spherical QD  of size  $R$,  surrounded by  a dielectric  (glasses,
polymers,  organic  solvents or  oxides  and  hydrides).   There is  a
hydrogenic impurity at the center of the QD.  In our model we vary the
shape  of   the  confinement,  namely  quasi-triangular   ($k  =  1$),
quasi-harmonic  ($k   =  2$)  and  quasi-square  ($k   \ge  10$)  (see
Eq.~(\ref{eq:pote})).   The right hand  side of  the figure  shows one
typical model  potential, corresponding  to the shape  index $k  = 80$
(quasi-square well). The  depth of the well is  $V_0$ and the diameter
is $2  R$.  The  neutral and negative  donor levels  are schematically
shown along with the  conduction band minimum (CBM) with corresponding
energies $E (D^0)$, $E (D^-)$ and $E_0$ respectively. Their respective
binding energies $E_B (D^0)$ and $E_B (D^-)$ are indicated by arrows.}
\end{figure*}

In order  to study  the shape dependence  of binding energy,  we model
\cite{ranj02,pand03} our external potential as follows:
\begin{eqnarray}
V_{ext} (r)  & =  & \left\{\begin{array}{lrl} (V_0/R^k)  r^k -  V_0 &&
\mbox{$r   \leq   R$}   \\   0   &&   \mbox{$r   >   R$}   \end{array}
\right. \label{eq:pote}
\end{eqnarray}
where $V_0$  is the depth of the  potential. This can be  given by the
conduction band  offset (valence band  offset) between the QD  and the
surrounding layer for the electron (hole). $R$ is the radius of the QD
and  $k$ assumes  positive  integral values  from  1 to  a very  large
number.  Changing the value of $k$  results in the change of the shape
of the potential.  In particular $k  = 1$ is quasi-triangular, $k = 2$
is  quasi-harmonic confinement and  $k \geq  10$ is  quasi-square well
confinement.   As  $k \rightarrow  \infty$,  the  potential becomes  a
square well.

The total ground state energy of  an N-electron system in terms of the
density is given as
\begin{eqnarray}
E [\rho]  & =  & T_S  [\rho] + E_H  [\rho] +  E_{xc} [\rho]  + E_{ext}
[\rho]+ E_{imp} [\rho] \label{eq:TE}
\end{eqnarray}
where   $T_S  [\rho]$   is  the   kinetic  energy   functional   of  a
non-interacting  many electron  system,  $E_H [\rho]$  is the  Hartree
energy, $E_{xc} [\rho]$ is the exchange-correlation energy functional,
$E_{ext}  [\rho]$  is  the  energy  functional  due  to  the  external
potential and $E_{imp} [\rho]$ is the energy functional because of the
impurity potential.   Note that $\rho  (r)$ is the charge  density. As
pointed  out   in  Sec.~\ref{s:intro}  we  have   used  two  different
approaches for  the calculation of  the exchange-correlation potential
($V_{xc}$), namely  the LDA in  its Gunnarsson-Lundqvist \cite{gunn76}
parameterized form and the exchange potential using HS scheme.  Within
the HS scheme, the exchange  potential is interpreted as the work done
in moving an electron in the field of its Fermi hole.

The HS  scheme calculates nearly exact exchange  interaction.  For the
atomic  case, the method  essentially reproduces  Hartree-Fock results
for all  the elements in  the periodic table \cite{sahn92}.   Also, HS
scheme  has   self-interaction  correction  built  in   it  and  hence
calculates  very good  eigenvalues  and total  energies  for a  single
electron system,  whereas the LDA is  known to yield  poor results for
the   hydrogen-like    atoms   due   to    non-cancellation   of   the
self-interaction of the electron.

A hydrogenic  impurity in  a host bulk  semiconductor gives rise  to a
shallow donor state  in the energy gap of the host.   We refer to this
as neutral donor state ($D^0$) and its energy as $E(D^0)$.  The energy
required to  take the electron from  the neutral donor  state into the
conduction band is termed as binding energy, $E_B (D^0)$ and is of the
order of 10  - 50 $meV$.  Hence the binding energy  of a neutral donor
can be written as
\begin{eqnarray}
E_B (D^0) = E_0 - E(D^0) \label{eq:be1}
\end{eqnarray}
where $E_0$ is the ground state energy of an electron in the conduction
band.

A  negative donor center  in a  semiconductor is  formed by  a neutral
donor  center trapping  an extra  electron.  The  binding energy  of a
negative donor can be interpreted as follows: We take an electron from
$D^-$ to the  conduction band. Initially the energy  of the system was
$E (D^-)$, the final energy is  $E_0 + E (D^0)$. Therefore the binding
energy  of a  negative donor  can be  given by  subtracting  the final
energy from the initial energy, \textit{i.e.}
\begin{eqnarray}
E_B (D^-) = E (D^0) + E_0 - E (D^-) \label{eq:be2}
\end{eqnarray}
where $E (D^0)$  is the ground state energy of a  neutral donor and $E
(D^-)$ is  the ground state  energy of a  negative donor. This  is the
energy required  to promote an  electron from $D^-$ to  the conduction
band.  Naturally  the binding  energy of a  negative donor  is smaller
than the neutral donor.

As  we  have mentioned  earlier  the  HS  scheme has  self-interaction
correction  built-in, hence  it calculates  very good  eigenvalues for
small  number of  electrons.  In  fact the  total energy  of  a single
electron with and  without the impurity potential is  exactly equal to
the eigenvalue of an electron  with and without the impurity potential
respectively.   In other  words, $E_0  =  \epsilon_0$ and  $E (D^0)  =
\epsilon_1$, where  $\epsilon_0$ is the  eigenvalue of an  electron in
the conduction band minimum and  $\epsilon_1$ is the eigenvalue of the
neutral donor. The expression for binding energy of a neutral donor in
terms of the eigenvalues is given as
\begin{eqnarray}
E_B (D^0) & = & \epsilon_0 - \epsilon_1 \label{eq:be3}
\end{eqnarray}
Thus the binding  energy of an electron in this case  is also equal to
the optical transition  energy.  We can also write  the binding energy
of a negative donor in terms  of the eigenvalues.  The total energy of
a hydrogen-like negative donor is given by the formula
\begin{eqnarray}
E (D^-)  & =  & 2 \epsilon_2  - \frac{1}{2} \int  \frac{\rho (\vec{r})
\rho (\vec{r^{\prime}})}{\left| \vec{r}  - \vec{r^{\prime}} \right|} d
\vec{r}  d \vec{r^{\prime}} +  E_{xc} [\rho]  - \int  V_{xc} (\vec{r})
\rho (\vec{r}) d \vec{r} \label{eq:TE1}
\end{eqnarray}
where $\epsilon_2$ is the eigenvalue of  two-electron system.  From
the   definition  of   the  binding   energy  of   a   negative  donor
(Eq.~(\ref{eq:be2}))
\begin{eqnarray}
E_B (D^-) & = & \epsilon_1 + \epsilon_{0} - 2 \epsilon_2 + \frac{1}{2}
\int  \frac{\rho (\vec{r})  \rho (\vec{r^{\prime}})}{\left|  \vec{r} -
\vec{r^{\prime}} \right|} d \vec{r} d \vec{r^{\prime}} - E_{xc} [\rho]
+ \int V_{xc} (\vec{r}) \rho (\vec{r}) d \vec{r}
\end{eqnarray}
\begin{eqnarray}
& = & \epsilon_1  + \epsilon_0 - 2 \epsilon_2 + U  - 2 E_{xc} [\rho] +
\int V_{xc} (\vec{r}) \rho (\vec{r}) d \vec{r} \label{eq:TE2}
\end{eqnarray}
where
\begin{eqnarray}
U    &   =    &   \frac{1}{2}    \int   \frac{\rho    (\vec{r})   \rho
(\vec{r^{\prime}})}{\left|  \vec{r}   -  \vec{r^{\prime}}  \right|}  d
\vec{r} d \vec{r^{\prime}} + E_{xc} [\rho]
\end{eqnarray}
is the effective electron-electron  interaction energy. We call it the
``Hubbard  $U$'' \cite{ranj02,  pand03}.  The  last two  terms  in the
Eq.~(\ref{eq:TE2})  cancel  each  other  (true only  for  one  orbital
system) in HS scheme but they do not cancel within LDA.  Hence, within
the HS  scheme, the  binding energy  of a negative  donor in  terms of
eigenvalues can be written as
\begin{eqnarray}
E_B (D^-) & = & \epsilon_1 + \epsilon_0 - 2 \epsilon_2 + U
\label{eq:be4}
\end{eqnarray}
Within the  LDA, $U =  U_{coul} + U_x  + U_{corr}$ whereas  within the
exchange only HS scheme, $U = U_{coul} + U_x$, where $U_{coul}$ is the
Coulomb energy,  $U_x$ is  the exchange energy  and $U_{corr}$  is the
correlation energy.
\section{Results And Discussion}
\subsection{\label{s:results1}The Binding Energy}
\begin{figure*}
\includegraphics{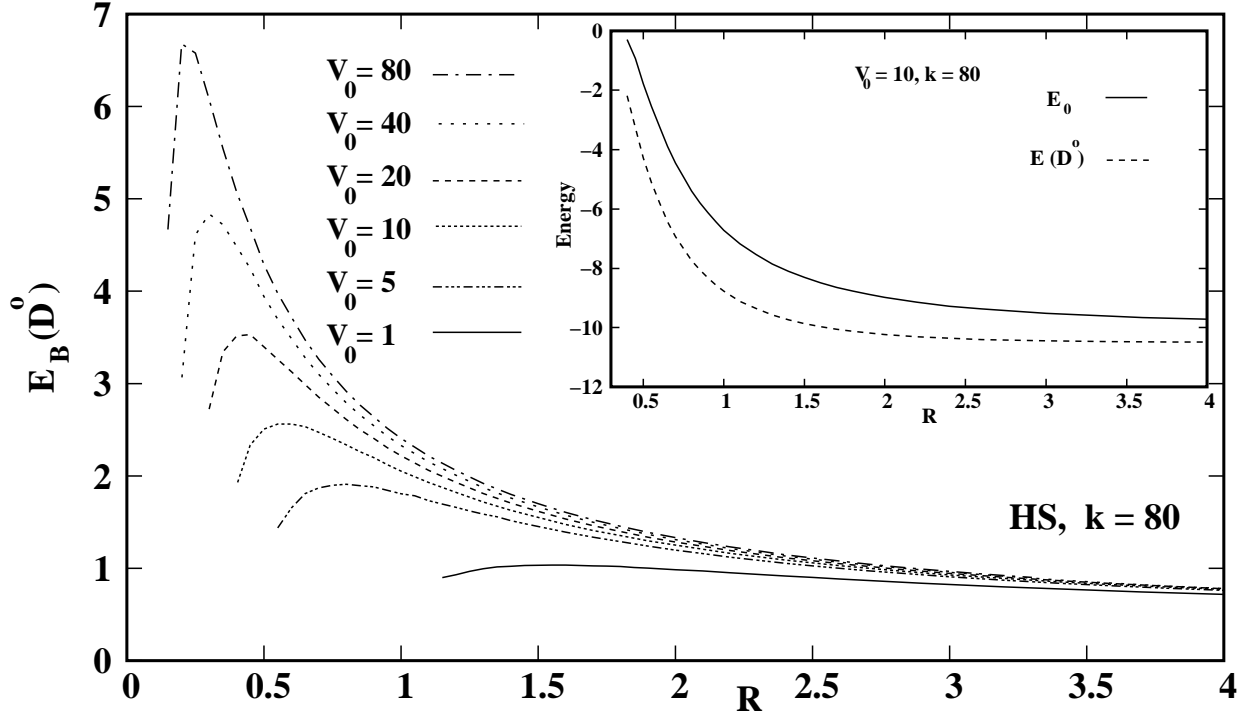}
\caption{\label{fig:msfig1}  The  binding energy  of  a neutral  donor
($E_B (D^0)$) is plotted as a function of the size $R$ of the QD.  The
shape  chosen is  quasi-square  well ($k  =  80$). The  depths of  the
potential well are  taken to be $V_0 =$  80, 40, 20, 10, 5  and 1. The
figure shows  a non-monotonic shallow  to deep (SHADES)  transition of
the binding  energy as the dot size  is reduced. In the  inset we plot
the conduction band  edge or LUMO (solid line)  and the impurity level
(dashed  line)  as a  function  of  size for  $V_0  =$  10  and $k  =$
80. Hartree units are employed.}
\end{figure*}
We present the  results of binding energy of  a hydrogen-like impurity
($Z =  1$).  As mentioned earlier  we are employing  Hartree units, in
which $\hbar = m_e  = e = 1$ and unit of energy  is 27.2 $eV$ and that
of length (size) is 0.53 \AA.  Unless otherwise stated all the results
presented are  based on HS  scheme. But the  HS scheme within  the one
electron system ($D^0$) is no  different from an electron moving in an
impurity potential;  essentially the Ranjan-Singh  (RS) \cite{ranj01}.
However  RS  scheme is  perturbative  and  therefore  its results  are
approximate. On the other hand,  our results are numerically exact. We
solve  the  one  -  dimensional  Schroedinger  equation  in  spherical
co-ordinates.

In Fig.~\ref{fig:msfig1} we depict the binding energy ($E_B (D^0)$) of
a neutral donor  ($D^0$) as a function  of the size $R$ of  the QD for
various different depths of  confinement potential, namely $V_0 =$ 80,
40, 20, 10, 5 and 1. The shape chosen is quasi-square well ($k = 80$).
The plots indicate  a monotonic increase in the  binding energy as the
size is reduced.  As we  decrease the size further, the binding energy
reaches a maximum and then it  decreases after a critical size. Thus a
non-monotonic shallow-deep  (SHADES) transition of the  donor level is
observed.   A  qualitative  explanation  of SHADES  transition  is  as
follows: The  binding energy is  the difference between the  energy of
conduction band  minimum (CBM)  and the impurity  energy.  As  the dot
size is decreased the CBM rises monotonically as $1/R^{\gamma}$, where
$\gamma \in  [1.2:2.0]$ \cite{sing00}.   On the other  hand, initially
the impurity  energy remains  relatively constant in  value as  $R$ is
decreased and then it increases rapidly.  This is because the impurity
wavefunction is localized and consequently the impurity charge density
is highly  confined. This charge  density does not sense  the boundary
initially.  We  have confirmed this picture by  a detailed examination
of the energies  and the wavefunctions. A typical  plot of CBM ($E_0$)
and the impurity ground state energy ($E (D^0)$) is shown in the inset
of Fig.~\ref{fig:msfig1}.  Even a  visual examination reveals that the
difference between the two energies is non-monotonic with size.  Below
a critical size,  which we define to be  $R_{SHADES}$, the donor level
becomes  shallow  again.   This  is because  below  $R_{SHADES}$,  the
kinetic  energy  expectation  value   increases  much  more  than  the
potential  energy expectation  value.   Note that  for  large $R$  the
binding energy approaches the bulk limit.

The binding energy is large for  large value of $V_0$. This is because
of  large confinement of  the charge  carrier.  We  also see  that the
critical   size  $R_{SHADES}$   at  which   SHADES   occurs  increases
monotonically  with  decrease  in   the  depth  of  the  well.   These
observations are recorded in Table~\ref{tab:table1}.  We explain these
as  follows:  For smaller  depth  of  the  confinement potential  (for
example   $V_0   =  1$),   the   strength   of   the  confinement   is
less. Conventional  textbooks define the strength of  the potential as
$V_0  R^2$.  Because  of  this, the  electronic  wavefunction is  more
spread out.   As one decreases  the size, the wavefunction  senses the
boundary much  before (\textit{i.e} at  larger size) than the  one for
which the  strength of  the confinement is  large (for example  $V_0 =
80$).   Hence for  small $V_0$,  the energy  of donor  level increases
rapidly. This implies that the binding energy is small for small $V_0$
and the  SHADES transition  occurs at larger  size. An  examination of
Table~\ref{tab:table1}    reveals   that    $V_0    R_{SHADES}^2$   is
approximately   constant.    In    other   words,   $R_{SHADES}   \sim
1/\sqrt{V_0}$.  Thus,  the binding energy  and its maximum  depends on
the  well  depth $V_0$.  The  latter  is  a surface  related  property
depending  on surface  termination of  dangling bonds,  the dielectric
coating, etc.   Thus, with experience,  it maybe possible  to engineer
the magnitude  of the binding energy, making  it shallow, intermediate
or deep depending on one's convenience. It suggests the possibility of
\textit{``defect engineering''} in QDs.

We pause to note that the hydrogenic impurity in bulk semiconductor is
quite  shallow.  The  carriers  in the  shallow  levels of  hydrogenic
impurities  are  easily promoted  into  the  conduction  band at  room
temperature,   thus  enhancing  the   conducting  properties   of  the
semiconductor  by several  orders of  magnitude. The  increase  in the
binding  energy  of  the  hydrogenic  impurity  with  decreasing  size
indicates  that  carrier {\it  ``freeze  out''}  will  occur.  Thus  a
nominally shallow donor will ``go deep''.

\begin{figure*}
\includegraphics{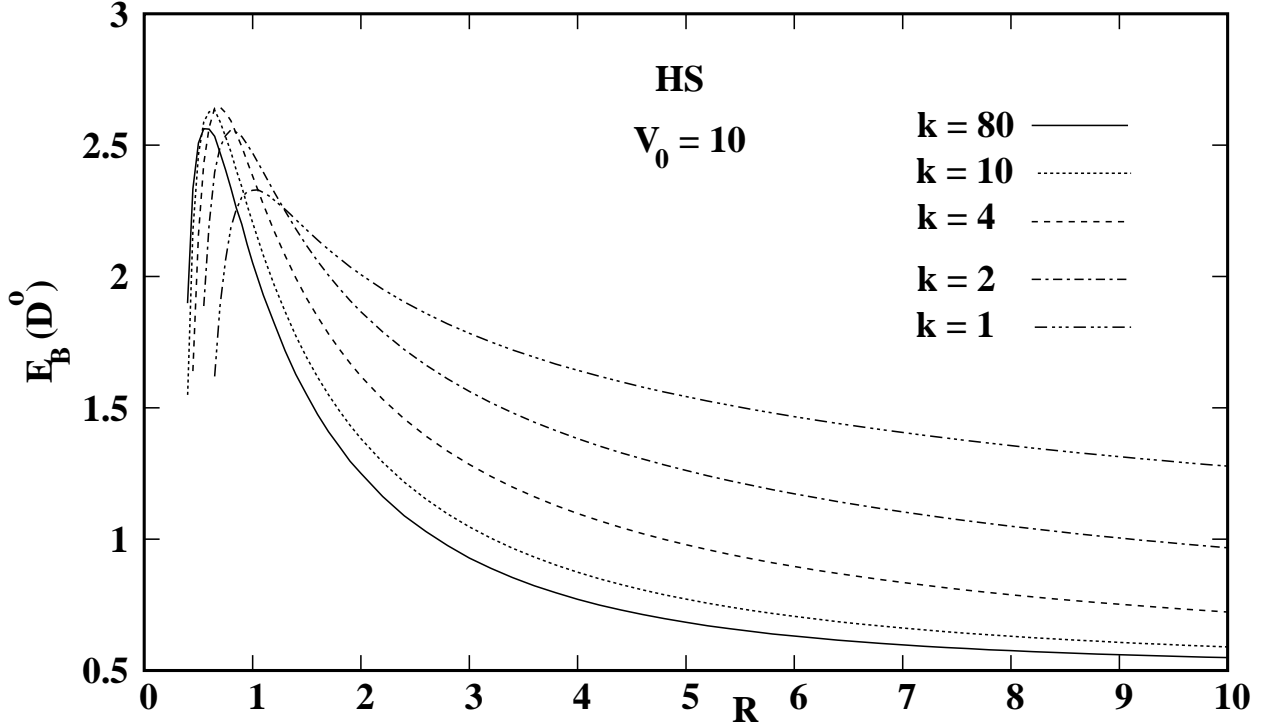}
\caption{\label{fig:msfig2}The  binding  energy  of  a  neutral  donor
($D^0$) is depicted as a function of  the size of the QD. The depth of
the well  is chosen to be  $V_0 = 10$  and the shape of  the confining
potential is varied, corresponding to the shape index $k =$ 80, 10, 4,
2 and 1. Hartree units are employed.}
\end{figure*}

In  Fig.~\ref{fig:msfig2} we  explore  the dependence  of the  binding
energy on  size as  the shape of  the confinement potential  is varied
from quasi-triangular  ($k =  1$) to quasi-harmonic  ($k = 2$)  and to
quasi-square ($k = 80$).  The potential  depth is kept fixed at $V_0 =
10$. We observe that, in  this case also, the binding energy increases
monotonically  as the  size is  reduced,  reaches a  maximum and  then
decreases. However, the critical dot size at which the maxima occur is
relatively  insensitive to  the  shape parameter  ($k$). Further,  the
dependence of the binding energy  maximum on the shape parameter ($k$)
is non-monotonic.   This is also  displayed in Table~\ref{tab:table2}.
We  point out  the possible  significance  of this  result.  When  the
charge distribution  is uniform throughout the spherical  QD, then the
shape of the potential  maybe taken to be quasi-harmonic.  If there is
excess charge  on the surface then  the shape of the  potential  maybe
taken  to be  quasi-square.  Thus,  depending on  how the  charges are
distributed in a  QD, it is possible to engineer  the magnitude of the
binding  energy.   Unlike  in  Fig.~\ref{fig:msfig1}, where  both  the
magnitude of  binding energy maxima and $R_{SHADES}$ are  sensitive to
the  depth of the  potential chosen,  we see  in Fig.~\ref{fig:msfig2}
that they are relatively less sensitive to the shape of the potential.

Notice from Fig.~\ref{fig:msfig2}  and the Table~\ref{tab:table2} that
the  binding energy  maxima  occur at  different  sizes for  $k =  80$
($R_{SHADES} =  0.55$) and $k =  10$ ($R_{SHADES} =  0.65$).  Here the
magnitude of the  binding energy maximum is larger for  $k = 10$ ($E_B
(D^0) = 2.63$) than for $k = 80$ ($E_B (D^0) = 2.56$).  Similarly, the
binding energy maxima  occur at slightly different sizes  for $k = 10$
($R_{SHADES}  = 0.65$)  and  $k =  4$  ($R_{SHADES} =  0.70$) but  the
magnitude  of the  binding energy  maxima are  roughly the  same ($E_B
(D^0) = $ 2.63 and 2.64 for $k =$ 10 and 4 respectively). Also, notice
that the binding energy maxima occur  at larger sizes as we change the
shape index $k$ from 4 to 2  and then 1, however, the magnitude of the
binding energy maxima decreases.

Figure~\ref{fig:msfig2}  also  reveals  that  the  binding  energy  is
largest when quasi-triangular confinement is used.  The binding energy
of  quasi-harmonic   confinement  is  larger   than  the  quasi-square
confinement.  This is because the charge carriers are more confined in
quasi-triangular   potential  than  quasi-harmonic   and  quasi-square
potential.    Similarly   charge  carriers   are   more  confined   in
quasi-harmonic than the quasi-square potential.  Thus one can tune the
magnitude of the binding energy by tuning the shape of the potential.

\begin{table}
\caption{\label{tab:table1}The table shows  the maximum binding energy
of $D^0$ and $D^-$ states  and the critical size $R_{SHADES}$ at which
the SHADES transition takes place with varied depth of the confinement
potential  ($V_0$).  We  note that  over  a large  range of  confining
potential  ($V_0 =$ 1  - 80),  the strength  of the  confinement ($V_0
R^2$)  at $R_{SHADES}$  remains relatively  unchanged (column  III and
VI).  The shape of  the confinement  potential chosen  is quasi-square
well ($k = 80$). Hartree units are employed.}
\begin{ruledtabular}
\begin{tabular}{ccccccc}
 $V_0$ & $R_{SHADES} (D^0)$ &  $V_0 R^2_{SHADES}(D^0)$ & $E_B (D^0)$ &
 $R_{SHADES} (D^-)$  & $V_0 R^2_{SHADES}(D^-)$ &  $E_B (D^-)$\\ \hline
 80 & 0.20 & 3.20  & 6.677 & 0.25 & 5.00 &1.449 \\ 40  & 0.30 & 3.60 &
 4.837 & 0.40 & 6.40 & 1.046 \\ 20 & 0.45 & 4.05 & 3.530 & 0.55 & 6.05
 & 0.740 \\ 10 & 0.55 & 3.03 & 2.562 & 0.80 & 6.40 & 0.536 \\ 5 & 0.80
 & 3.20 &  1.911 & 1.05 &  5.51 & 0.396 \\ 1  & 1.60 & 2.56  & 1.035 &
 2.10 & 4.41 & 0.184 \\
\end{tabular}
\end{ruledtabular}
\end{table}

\begin{figure*}
\includegraphics{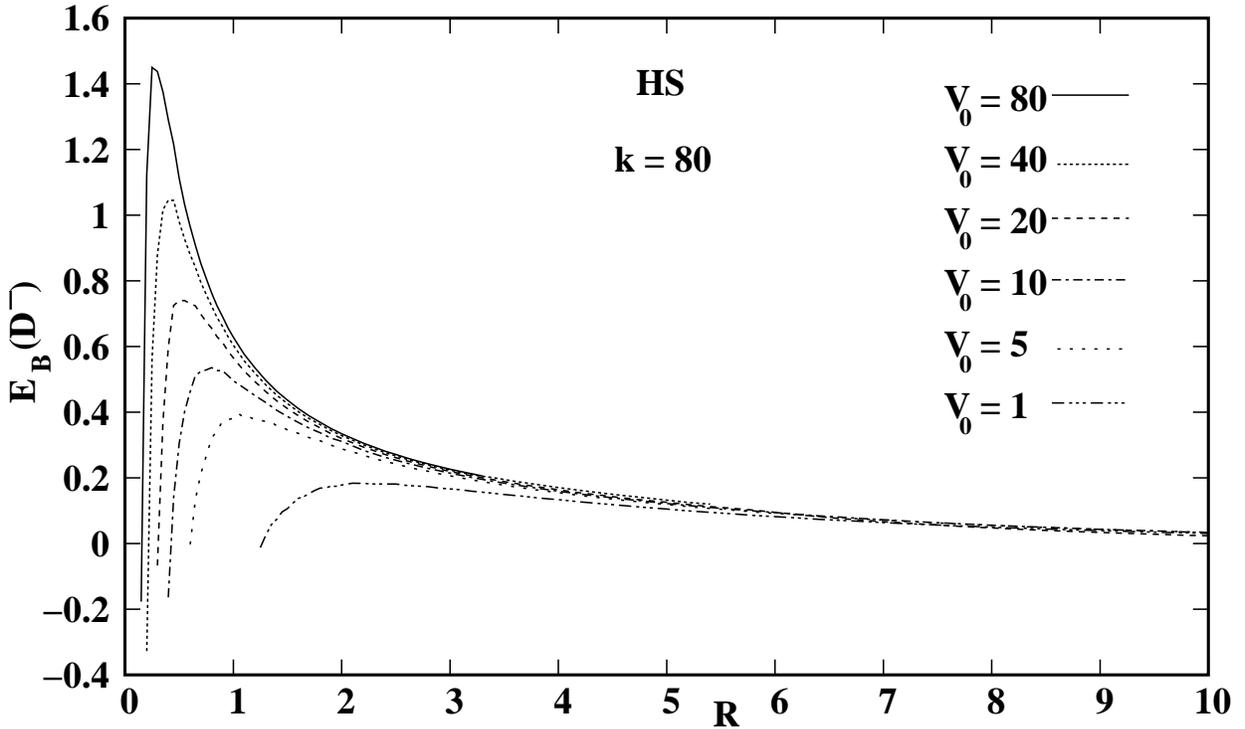}
\caption{\label{fig:msfig3}  The binding  energy of  a  negative donor
($D^-$) is  plotted as  a function of  the size  of the QD.  The shape
chosen  is  quasi-square  well ($k  =  80$).   The  depths of  of  the
potential well are chosen  to be $V_0 =$ 80, 40, 20,  10, 5 and 1. The
figure shows  a a non-monotonic  SHADES transition. Hartree  units are
employed.}
\end{figure*}
Figure~\ref{fig:msfig3} shows  the binding energy of  a negative donor
($D^-$) as a function of the  size of the QD.  The parmaeters used are
the  same as  for  Fig.~\ref{fig:msfig1}. Once  again a  non-monotonic
SHADES transition is observed.  For very small size the binding energy
becomes negative.   This implies that  the impurity level  is resonant
with the conduction band.  Note, SHADES occurs at larger size when the
depth of the potential is taken  to be smaller.  Finally at large size
the binding energy approaches the bulk limit.

If we compare the binding energy plot of $D^0$ (Fig.~\ref{fig:msfig1})
with the binding energy plot of $D^-$ (Fig.~\ref{fig:msfig3}), we note
the following: The binding energy of $D^-$ is smaller than $D^0$ for a
fixed depth  of the  confinement $V_0$ and  shape index $k$.   This is
obvious because the  $D^-$ level is higher than the  $D^0$ level by an
amount approximately  equal to $U$, the effective  electron - electron
interaction energy.  For a fixed  $k$ and $V_0$, the SHADES transition
for  $D^-$  occurs  at   larger  size  than  for  $D^0$  \textit{i.e.}
$R_{SHADES}  (D^-)   >  R_{SHADES}   (D^0)$.   This  is   because  the
wavefunction  of the  negative donor  is more  spread out  due  to the
Coulomb repulsion  whereas, the wavefunction of neutral  donor is more
confined.   As the size  becomes small,  the wavefunction  of negative
donor senses the boundary relatively earlier and gets squeezed.  As we
decrease the size further, below a critical size ($R_{SHADES} (D^-)$),
the charge  carriers are delocalized and  hence leak out  of the well.
Therefore the binding energy of  negative donor ($D^-$) decreases at a
size ($R_{SHADES}(D^-)$) larger than  the size ($R_{SHADES} (D^0)$) at
which the binding energy of neutral donor ($D^0$) decreases.

\begin{figure*}
\includegraphics{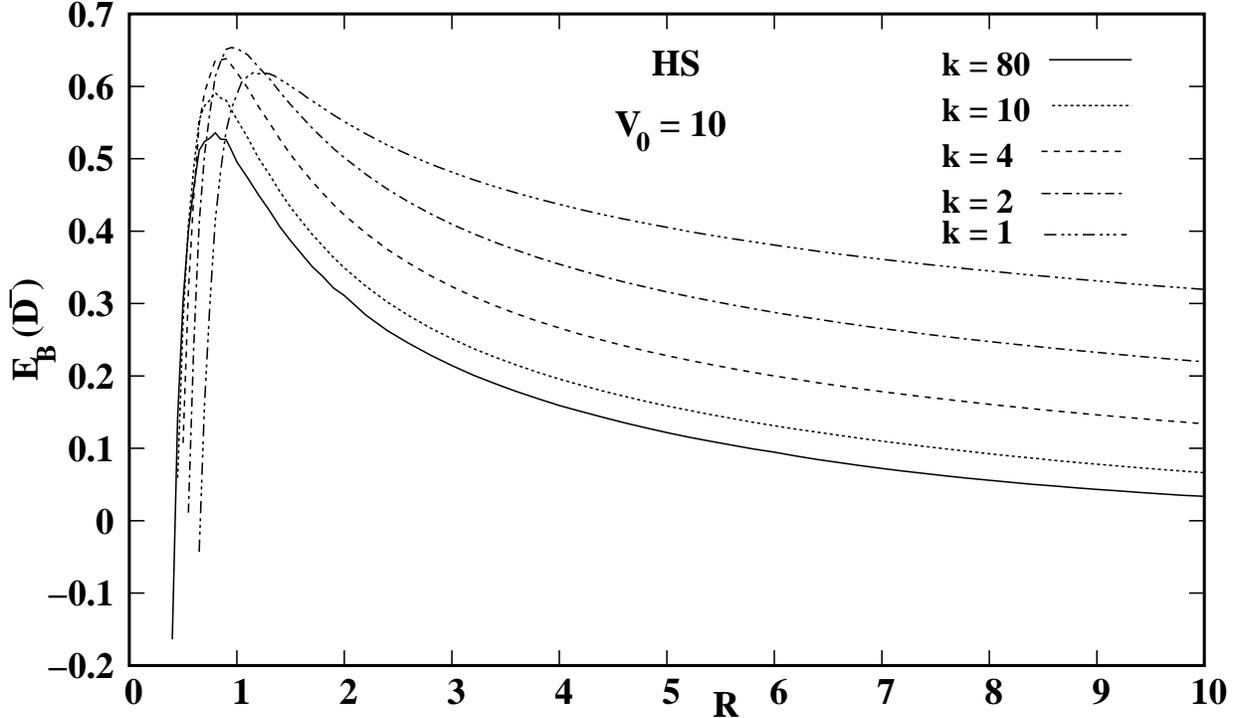}
\caption{\label{fig:msfig4}The  binding  energy  of a  negative  donor
($D^-$) is depicted as a function of  the size of the QD. The depth of
the well  is chosen to be  $V_0 = 10$  and the shape of  the confining
potential is varied, corresponding to the shape index $k =$ 80, 10, 4,
2 and 1. Hartree units are used.}
\end{figure*}

We depict  the dependence of  the negative donor  on the shape  of the
confinement potential  in Fig.~\ref{fig:msfig4}.  The  parameters used
are  the  same  as  for  Fig.~\ref{fig:msfig2}.   The  binding  energy
exhibits a  non-monotonic SHADES transition. The critical  dot size at
which the maxima occur is even more insensitive to the shape parameter
($k$)  than  it  is  for   the  neutral  donor,  as  is  evident  from
Fig.~\ref{fig:msfig4}  and  a  comparison  of  columns II  and  IV  of
Table~\ref{tab:table2}.  The dependence  of the binding energy maximum
on the shape parameter ($k$) is non-monotonic as it is for the neutral
donor.  The  binding energy maxima occur  at larger size  as we change
the shape index $k$ from 2  to 1, however the magnitude of the binding
energy  decreases.    These  and   other  aspects  are   displayed  in
Table~\ref{tab:table2}.  Note the binding energy of the negative donor
is much smaller than the binding energy of the neutral donor.
\begin{table}
\caption{\label{tab:table2}The table shows  the maximum binding energy
of $D^0$ and $D^-$ and $R_{SHADES}$ at with varied shapes ($k$) of the
confinement potential.  The depth  of the confinement potential chosen
is $V_0 = 10$. Hartree units are employed.}
\begin{ruledtabular}
\begin{tabular}{ccccc}
 $k$ &  $R_{SHADES} (D^0)$ & $E_B  (D^0)$ & $R_{SHADES}  (D^-)$ & $E_B
 (D^-)$\\ \hline 80 & 0.55 & 2.562 & 0.80 & 0.536 \\ 10 & 0.65 & 2.633
 & 0.80 & 0.591 \\ 4 & 0.70 & 2.645 & 0.90 & 0.638 \\ 2 & 0.80 & 2.558
 & 0.95 & 0.654 \\ 1 & 1.05 & 2.329 & 1.15 & 0.619 \\
\end{tabular}
\end{ruledtabular}
\end{table}

We have also  calculated the binding energies of  neutral and negative
donors using Eq.~(\ref{eq:be3}) and Eq.~(\ref{eq:be4}). As pointed out
earlier, we  use two  different methodologies, LDA  and HS  scheme. We
have compared the binding energies  of a neutral donor, obtained using
Eqs.~(\ref{eq:be1})  and (\ref{eq:be3}).   We have  also  compared the
binding    energies   of    a   negative    donor,    obtained   using
Eqs.~(\ref{eq:be2}) and (\ref{eq:be4}), within the HS scheme.  We find
that  the  HS scheme  gives  identical  values  of $E_B  (D^0)$  using
Eqs.~(\ref{eq:be1})  and  (\ref{eq:be3}).   It  also  gives  identical
values of $E_B (D^-)$ using Eqs.~(\ref{eq:be2}) and (\ref{eq:be4}). We
did the same  exercise within the LDA. But LDA  gives poor results for
both cases.
\subsection{\label{s:results2}The Optical Gap}
\begin{figure*}
\includegraphics{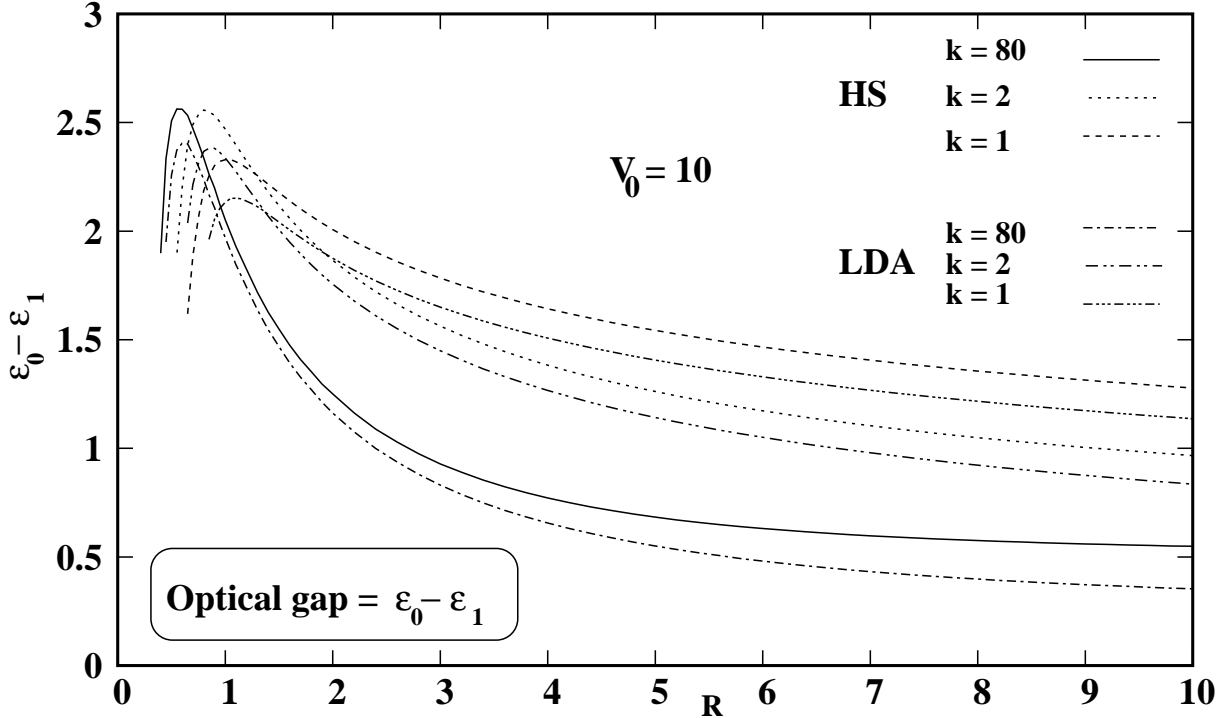}
\caption{\label{fig:msfig5}The eigenvalue difference of an electron in
the  conduction band  minimum  and the  neutral  donor ($\epsilon_0  -
\epsilon_1$) is plotted as a function  of the size of QD. The depth of
the potential  is chosen to be  $V_0 = 10$. Three  different shapes of
the  potential are  chosen corresponding  to $k  =$ 1, 2  and  80. For
comparison we plot results of the  calculations of both LDA and the HS
scheme. Hartree units are employed.}
\end{figure*}

Figure~\ref{fig:msfig5} shows the eigenvalue difference ($\epsilon_0 -
\epsilon_1$) between  an electron in  the conduction band  minimum and
the   neutral   donor.   This   difference   ($\Delta  \epsilon$)   is
representative of  the optical gap  \cite{jana78} of the  system.  The
depth of the potential chosen is $V_0$ = 10. Three different shapes of
the confinement potential are chosen,  corresponding to $k =$ 1, 2 and
80. We show results obtained using both  the LDA and the HS scheme. We
observe  that   the  optical  gap  also   shows  non-monotonic  SHADES
transition.  As noticed earlier, the optical gap obtained using the HS
scheme is identically equal to  the binding energy of a neutral donor,
$E_B  (D^0)$ (Eq.~(\ref{eq:be1})).   Thus  within the  HS scheme,  the
optical gap follows  the same track with the size  ($R$) and the shape
parameter $k$ as shown  for $E_B (D^0)$ in Fig.~\ref{fig:msfig2}.  The
results obtained using LDA are different from those obtained using the
HS scheme.  The  figure reveals that the optical  gap calculated using
LDA is smaller than that of  the HS scheme.  This is because magnitude
of eigenvalue within the LDA  for hydrogen-like systems is smaller due
to the self-interaction of electron.

\begin{figure*}
\includegraphics{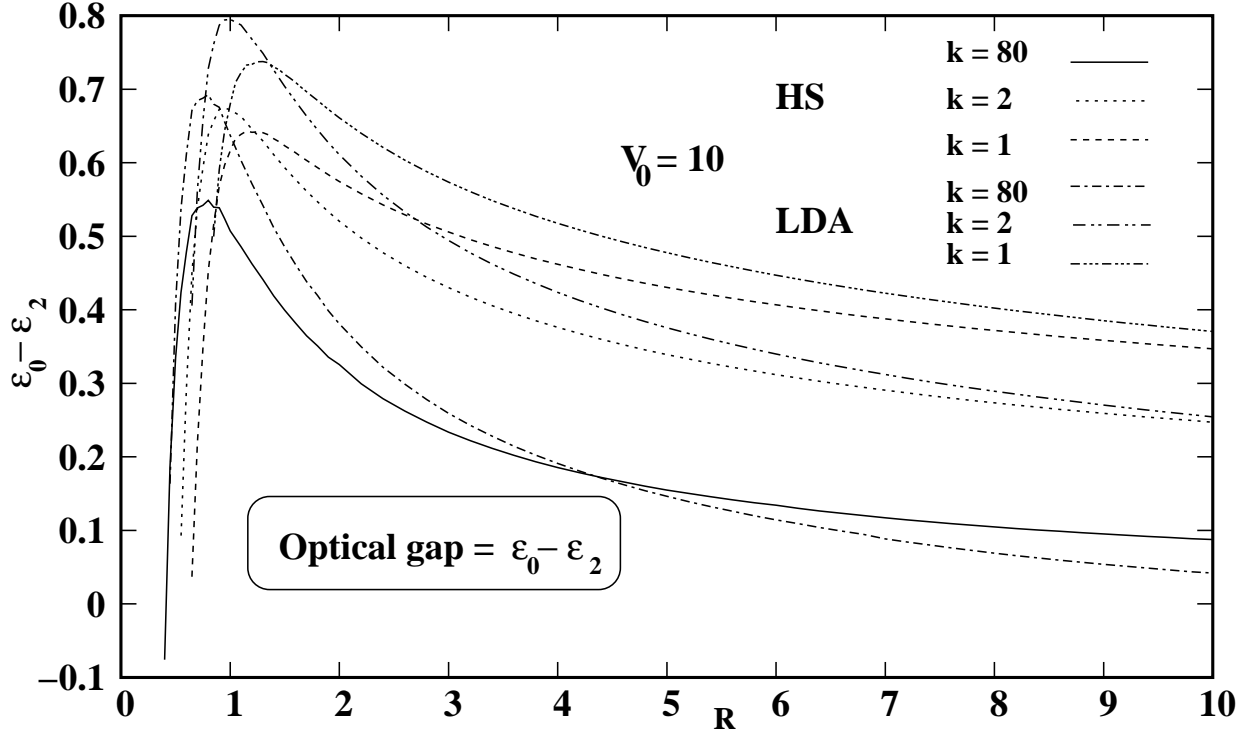}
\caption{\label{fig:msfig6}The eigenvalue difference of an electron in
the  conduction band  minimum and  the negative  donor  ($\epsilon_0 -
\epsilon_2$) is plotted as a function  of the size of QD. The depth of
the potential  is chosen to be  $V_0 = 10$. Three  different shapes of
the  potential are  chosen corresponding to $k  =$ 1, 2  and  80.  For
comparison we plot results of the  calculations of both LDA and the HS
scheme. Hartree units are employed.}
\end{figure*}

Figure~\ref{fig:msfig6} shows the eigenvalue difference ($\epsilon_0 -
\epsilon_2$) between  an electron in  the conduction band  minimum and
the negative donor  (optical gap). The parameters chosen  are the same
as for Fig.~\ref{fig:msfig5}. For  comparison we show results obtained
using  both the  LDA and  the HS  scheme.  Once  again we  observe the
non-monotonic SHADES  transition in the  optical gap. The  optical gap
shows  similar features  with size  and the  shape of  the confinement
potential as those shown in Fig.~\ref{fig:msfig4}.  In contrast to the
single electron  case, we find  that the optical gap  calculated using
LDA is larger than that of the HS scheme.

In the  context of  the optical  gap, we would  like to  elucidate the
connection between our calculation and the work of Bhargava \textit{et
al.}\cite{bhar94}. The deep defect levels in QD would act as efficient
traps for photo-excited  carriers.  Enhanced luminescence results when
these  carriers undergo  radiative recombination.   This  scenario was
proposed by Bhargava \textit{et  al.}  who found that incorporating Mn
impurities   in  ZnS   nanocrystallites  results   in   a  spectacular
enhancement  of   the  luminescence   efficiency  ($18  \%$)   with  a
corresponding    decrease   of    lifetime   from    milliseconds   to
nanoseconds. Thus high luminescence is perhaps at least partly related
to the deepening  of impurity levels.  The deepening  of the nominally
shallow hydrogenic level observed by  us implies that we maybe able to
use them as efficient luminescence center.

\begin{figure*}
\includegraphics{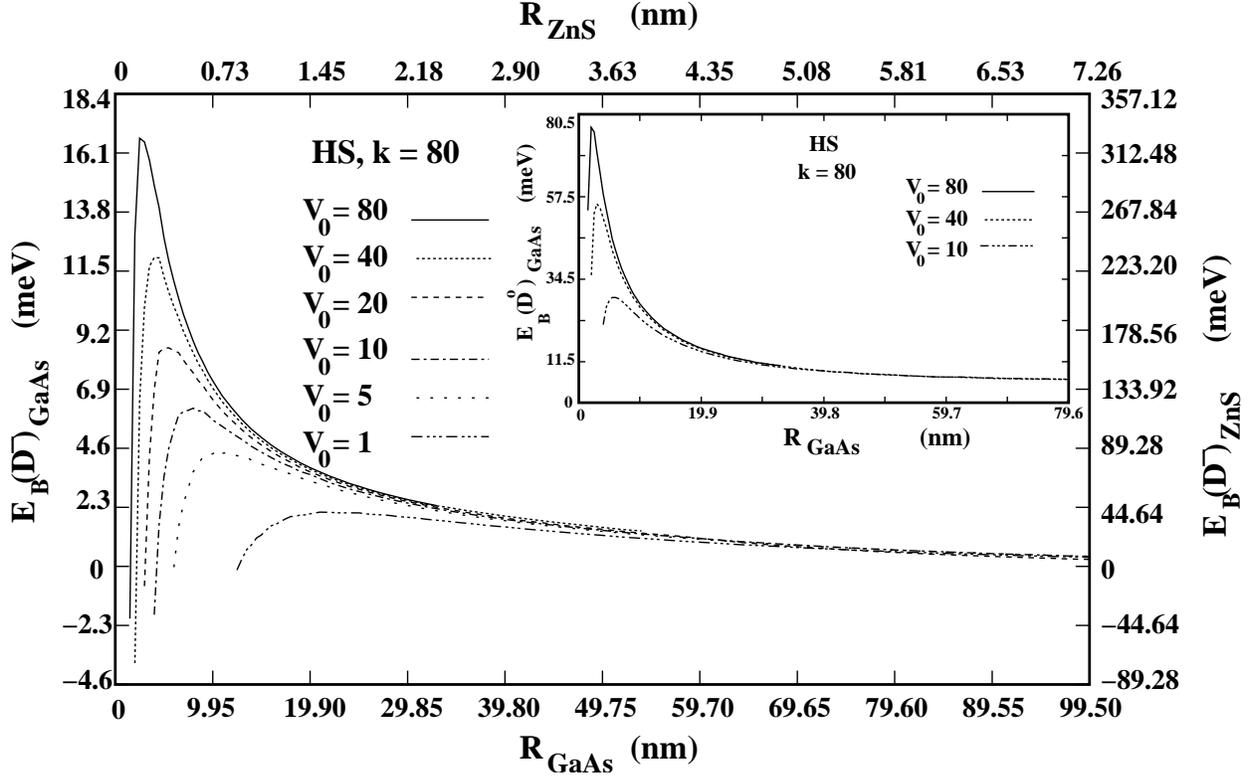}
\caption{\label{fig:msfig7}This  plot  shows  the  dependence  of  the
binding energy  of negative  donor (\textit{meV}) on  the QD  size $R$
(\textit{nm}) and  the potential well  depth $V_0$.  The shape  of the
potential chosen is quasi-square well ($k = 80$). The materials chosen
are GaAS and  ZnS. Note that for ZnS, the defect  levels are very deep
(0.2 - 0.3 $eV$).  Hence carrier \textit{\textbf{``freeze out''}} is a
distinct possibility in ZnS. Also,  note that the SHADES occurs at very
small  size ($R  <$ 1  $nm$  ) in  ZnS nanocrystals  whereas, in  GaAs
nanocrystals SHADES occurs at $R \simeq$ 10 $nm$.}
\end{figure*}

\section{Conclusion}
\label{s:conclusion}
Shallow dopants play a  critical role in semiconductor technology. The
present   work    as   encapsulated   in    Figs.~\ref{fig:msfig1}   -
\ref{fig:msfig4} presents  an interesting scenario.   This scenario is
presented for  the technologically relevant  case GaAs and ZnS  QDs in
Fig.~\ref{fig:msfig7}.   Figure~\ref{fig:msfig7} is  for  a negatively
charged donor where many-body effects  are relevant.  The inset is for
neutral  donor \cite{scal}.  However  as noted  earlier both  $D^0$ or
$D^-$  show similar  behaviour. A  nominally shallow  donor  ($D^0$ or
$D^-$) becomes deep  as the dot size is  decreased.  This implies that
carriers will  \textit{\textbf{``freeze out''}}.  On  further decrease
of the dot size the dopants may once again become shallow (SHADES). An
examination of  Fig.~\ref{fig:msfig7} also indicates  that the binding
energy and its maxima depends on  the well depth $V_0$.  The latter is
a  surface  related  property  depending  on  surface  termination  of
dangling bonds,  the dielectric coating, etc.   Thus, with experience,
it maybe  possible to  engineer the magnitude  of the  binding energy,
making   it  shallow,   intermediate  or   deep  depending   on  one's
convenience.  This suggests  the so called ``synthetic tailorability''
\cite{heat03} of  binding energy by  selecting a suitable dot  size or
dielectric   coating.   This   may   usher  in   the  possibility   of
\textit{``defect   engineering''}   in   QDs.    As  is   clear   from
Fig.~\ref{fig:msfig7}, the doped GaAs QDs of size $R \le 20$ $nm$ (200
\AA)   will  be   susceptible   to  carrier   \textit{\textbf{``freeze
out''}}. Similarly the  doped ZnS QDs of size $R \le  2$ $nm$ (20 \AA)
will be susceptible to carrier \textit{\textbf{``freeze out''}}.

We also observed the SHADES transition in the optical gap. We note the
relevance  of  our calculations  in  Bhargava  \textit{et al.}'s  work
\cite{bhar94}.  In  this work, they  proposed that the deep  levels in
materials  such as  Mn  doped ZnS  would  act as  efficient traps  for
photo-excited  carriers.   This leads  to  enhanced luminescence  when
these  carriers undergo  radiative  recombination. If  we accept  this
point of view,  then a suitably tailored ``deep'' dopant  may act as a
luminescent center.

We note that we are working in EMT-LDA approximation. Preliminary work
on an  elaborate tight binding  calculation on neutral donor  seems to
indicate SHADES  behaviour \cite{sing98}. The donor  impurities in the
present work are located at the  center of a spherical quantum dot. It
would be interesting to carry out investigations in which the shape of
the QD is varied and the donor  is moved off center.  It would also be
of interest to  investigate helium-like donors, e.g. S  in Si. In this
case many-body effects  would be significant, since one  would have at
least three charge  states of the donor, namely  $D^+, D^0$ and $D^-$.
We plan to undertake some of these tasks in future.

\begin{table}
\caption{\label{tab:table3}This  table shows  the potential  depths in
$eV$ for materials, namely GaAS and ZnS.}
\begin{ruledtabular}
\begin{tabular}{ccc}
$V_0$ &  $V_0$ (GaAs) in  $eV$ &  $V_0$ (ZnS) in  $eV$ \\ \hline  80 &
0.920 & 17.86 \\ 40 & 0.460 & 8.93  \\ 20 & 0.230 & 4.46 \\ 10 & 0.115
& 2.23 \\ 5 & 0.058 & 1.12 \\ 1 & 0.012 & 0.22 \\
\end{tabular}
\end{ruledtabular}
\end{table}
\section*{Acknowledgement}
This work was supported by the Department of Atomic Energy through the
Board    of    Research   in    Nuclear    Sciences,   India    (Grant
No. 2001/37/16/BRNS).


\begin{thebibliography}{10}
\bibitem{scal}  The size  in  $nm$  and binding  energy  in $meV$  are
converted as follows:  Size $R$ in $nm$ is  obtained from the relation
$R  ~(nm) =  0.053 ~\epsilon  ~R/m^*$ and  binding energy  in  $eV$ is
obtained from  the relation $E_B  ~(eV) = 27.2  ~m^* ~E_B/\epsilon^2$.
For GaAs, $m^*  = 0.067$ and $\epsilon = 12.58$.   This gives one unit
of size and binding energy for  GaAs equal to 9.95 $nm$ and 11.5 $meV$
respectively.  Thus,  $R ~(nm)  =  9.95 ~R$  and  $E_B  ~(meV) =  11.5
~E_B$. For ZnS, $m^* = 0.65$ and $\epsilon = 8.9$. This gives one unit
of size and binding energy for ZnS equal to 0.726 $nm$ and 223.2 $meV$
respectively.  Thus, $R  ~(nm) =  0.726 ~R$  and $E_B  ~(meV)  = 223.2
~E_B$.
\end{thebibliography}
\end{document}